\documentstyle[preprint,prb,aps]{revtex} 
\tighten  
\begin{document}  
\draft  
\title{Electronic structure, phase stability and chemical bonding in Th$_2$Al and Th$_2$AlH$_4$}
\author{P. Vajeeston$^{1*}$, R. Vidya$^1$, P. Ravindran$^1$, H. Fjellv{\aa}g$^{1,2}$, 
A. Kjekshus$^1$ and A. Skjeltorp$^2$}
\address{$^1$Department of Chemistry, University of Oslo, Box 1033, Blindern, N-0315, Oslo, Norway.}
\address{$^2$ Institute of Energy Technology, P.O.Box 40, Kjeller, N-2007, Norway. } 
\date{\today} 
\maketitle
\begin{abstract}
{\sloppy } 
We present the results of theoretical investigation on the electronic
structure, bonding nature and ground state properties of Th$_2$Al and
Th$_2$AlH$_4$ using generalized-gradient-corrected first-principles 
full-potential density-functional calculations. Th$_2$AlH$_4$ has
been reported to violate the "2\,\AA rule" of H-H separation in hydrides.
From our total energy as well as force-minimization calculations, we found
a shortest H-H separation of 1.95\,{\AA} in accordance with recent high resolution
powder neutron diffraction experiments. When the Th$_2$Al matrix is
hydrogenated, the volume expansion is highly anisotropic, which is quite
opposite to other hydrides having the same crystal structure. The bonding
nature of these materials are analyzed from the density of states, 
crystal-orbital Hamiltonian population and valence-charge-density analyses.
Our calculation predicts different nature of bonding for the H atoms
along $a$ and $c$. The strongest bonding in Th$_2$AlH$_4$ is between 
Th and H along $c$ which form dumb-bell shaped H-Th-H subunits. 
Due to this strong covalent interaction there is very small amount
of electrons present between H atoms along $c$ which makes
repulsive interaction between the H atoms smaller and this is the precise
reason why the 2\,{\AA} rule is violated. The large difference in
the interatomic distances between the interstitial
region where one can accommodate H in the $ac$ and $ab$ planes along
with the strong covalent interaction between Th and H 
are the main reasons for highly anisotropic volume
expansion on hydrogenation of Th$_2$Al.

\end{abstract}
\pacs{PACS numbers: 71., 81.05.Je, 71.15.Nc, 71.20.-b}

\section{INTRODUCTION}
Hydrides of intermetallics have been extensively studied because of their
applications in re-chargeable batteries. Unfortunately, most metals that
absorb large amounts of hydrogen are either heavy or expensive.\cite{rao85}
Consequently, there is a constant search for hydrides that may be suitable
for practical applications. First of all, it is very important to
understand how crystal structural evolution takes place in the course of
hydrogenation. Numerous studies have been done to explain
observed stabilities, stoichiometries, and preferred H sites in
hydrides of metallic and intermetallic compounds. Structural studies of
hydrides have provided empirical rules\cite{westlake84} that can be used to
predict the stability of the H sublattice in a given metal configuration. A 
survey of stable hydrides show that the H$-$H distance does not go below
2.1\,{\AA} (the 2\,{\AA} rule) with a minimum radius of   
0.4\,{\AA} for the inter-site to be used for the accommodation of H. 
These rules have been used to predict new hydrides whose
existence is verified experimentally.\cite{rao85,westlake84,switendick79}
\par
The review of Yvon and Fischer\cite{yvon88} states that 
Th$_2$AlH$_4$\cite{bergsma61} and K$_2$ReH$_9$\cite{yvon88,abrahams64} 
are violate the 2\,{\AA} rule, the shortest H$-$H separation being
1.79, and 1.87, respectively. K$_2$ReH$_9$ is classified among complex
transition metal hydrides, which comprise highly covalent solids with nonmetallic
properties. Th$_2$AlH$_4$, on the other hand, has metallic character.  
\par
Th$_2$Al\cite{havinga72} together with  Zr$_2$Fe, Zr$_2$Co and Zr$_2$Ni  
crystallize in the CuAl$_2$-type structure, whereas their
hydrides form rather different crystal structures. Zr$_2$Fe and Zr$_2$Co
form the isostructural deuterides Zr$_2$MD$_5$ (M = Fe,Co)\cite{yartys98} 
with a change in symmetry from ${I4/mcm}$ to ${P4/ncc}$ on deuteration.  
Th$_2$AlH$_4$\cite{bergsma61,sorby20} and Zr$_2$NiH$_{4.74}$\cite{chikdene89} 
are formed without any change in the symmetry from their parent structures. 
Th$_2$AlH$_4$ belongs to the exclusive class which does not obey the 2\,{\AA} rule.
The lattice expansion along $a$ and $c$ has proved to be highly 
anisotropic on hydrogenation of Th$_2$Al. In order to shed light on this
effect we need theoretical understanding about bonding
nature in this compound. Further, the understanding of the lattice
expansion and distortion during hydrogenation will be important for the
evaluation of stability of the hydride. So, we have made detailed study
of Th$_2$Al and Th$_2$AlH$_4$ by first-principle calculations.
\par
Two different powder neutron diffraction (PND) of Th$_2$AlH$_4$ give 
different H-H separations,${viz.}$ the older\cite{bergsma61} value is 1.79\,{\AA} and  
the more recent\cite{sorby20} value is 1.97\,{\AA}. So one aim of this 
study has been to solve this discrepancy.
In principle, the stability of hydrides can be evaluated directly from a
theoretical study of the total energy. However, owing to the complexity 
of the structure of transition metal hydrides, no reliable theoretical
heat of formation has hitherto been reported.\cite{nakamura98} 
Nakamura {\it et al.}\cite{nakamura98} were the first to calculate heat
of formation. However, these authors obtained positive and unrealistically 
large heat of formation even for stable La-Ni based hydrides except for
(La$_2$Ni$_{10}$H$_{14}$).\cite{murray81,bowerman79,hubbard83} 
This unfavorable result clearly indicates that local relaxation
of the metal atoms surrounding the hydrogens must be 
included in the calculations in order to predict the structural 
stability parameters. Hence our calculations take into account local
relaxation by optimizing the atom positions globally.
\par
We present the electronic structure of Th$_2$Al and Th$_2$AlH$_4$, obtained by
the full-potential linearized-augmented plane wave (FPLAPW) method. A
central feature of the paper is the evaluation of the electronic structure and
bonding characteristics on introduction of H into the Th$_2$Al matrix. In addition
to regular band-structure data, we also provide crystal
orbital Hamiltonian population (COHP)\cite{boucher98,dronskowski93} results to
illustrate the chemical bonding in more detail.    
\par
This paper is organized as follows. Details about the involved structure and
computational method are described in Sec.~\ref{sec:cry}.  
Sec.~\ref{sec:r&d} gives the results of the calculations and comparisons
with the experimental findings. Conclusions are briefly summarized in Sec.~\ref{sec:s&c}.

\section{Structural Details} 
\label{sec:cry}
Th$_2$Al and Th$_2$AlH$_4$ crystallize in space group ${I4/mcm}$ with the 
lattice parameters $a$ = 7.618, $c$ = 5.862\,{\AA}\cite{pearson86} for
Th$_2$Al and $a$ = 7.626,$c$ = 6.515\,{\AA} for Th$_2$AlH$_4$\cite{sorby20}
The crystal structure of Th$_2$AlH$_4$ is illustrated in Fig.\,\ref{fig:cry}.  
The crystal structure of Th$_2$Al contains four crystallographically different
interstitial sites, which are the suitable sites for hydrogen accommodation,  
$16l$ and $4b$ each coordinated to four Th, $32m$ coordinate to three
Th and one Al $16k$ coordinate to two Th and two Al.
Each $16l$ based intersite tetrahedra share a common face with
another $16l$-based tetrahedron, whereas the $4b$-based tetrahedron
shares each of its four faces with $16l$-based tetrahedra. Some of
the tetrahedral intersites are closely separated owing to the face sharing of
the coordination polyhedra. According to the experimental 
findings,\cite{bergsma61,sorby20} the $16l$ sites are fully 
occupied in Th$_2$AlD$_4$, and also the structure is 
completely ordered.

\subsection{ Computational details} 
In our calculations we use the full-potential linearized augmented 
plane wave (FP-LAPW) method in a scalar relativistic version without 
spin-orbit coupling as embodied in the WIEN97 code.\cite{blaha97} In brief,
this is an implementation of density-functional theory (DFT) with different 
possible approximations for the exchange and correlation potential, including
the generalized-gradient approximation (GGA).
The Kohn-Sham equations are solved using a basis of linearized augmented plane
waves.\cite{singh94} For the exchange and correlation potential, we used the
Perdew and Wang\cite{perdew96} implementation of GGA. For the potential
and charge density representations, inside the muffin-tin spheres the wave 
function are expanded in spherical harmonics with $l_{max}$ = 10, and non 
spherical components of the density and potential are included up to $l_{max}$ = 6.
In the interstitial region they are represented by Fourier series and thus they 
are completely general so that such a scheme is termed full-potential 
calculation. The present calculations we used muffin-tin radii of 2.5, 
2.1 and 0.9 Bohr for Th, Al and H respectively.
\par
The  basis set includes  7$s$, 7$p$, 6$d$ and 5$f$ valence and 6$s$ and 6$p$ semi-core 
states for Th, 3$s$, 3$p$ valence and 2$p$ semi-core states for Al and 1$s$ states 
for H. These basis functions were supplemented with local orbitals\cite{dsingh} 
for additional flexibility to the representation of the semi-core states and
for generalization of the linearization errors. We have included the local orbitals
for Th $6s$, $6p$ and Al $2p$ semicore states. 
In all our calculations we have used tetrahedron method on a grid of 
102 {\bf k} points in the irreducible part of the hexagonal Brillouin 
zone (IBZ),\cite{blochl94} which
corresponds to 1000 {\bf k} points in the whole Brillouin zone. The
calculations are done at several cell volumes (around the equilibrium volume)
for both Th$_2$Al and Th$_2$AlH$_4$ and corresponding total energies are evaluated
self-consistently by iteration to an accuracy of 10$^{-6}$ Ry./cell.
Similar densities of {\bf k} points were used for the force minimization
and ${c/a}$ optimization calculations.
\par
In order to measure the bond strengths we have computed the COHP\cite{dronskowski93}
 which is adopted in the TBLMTO-47 package.\cite{andersen75,cohp} COHP is 
the density of states weighted by the corresponding Hamiltonian matrix elements, 
which if negative indicates a bonding character and positive indicates an
anti-bonding character. The simplest way to investigate the bonding between
two interacting atoms in the solid would be to look at the complete COHP 
between them, taking all valence orbitals into account. However, it may
sometimes be useful to focus on pair contributions of some specific orbitals.

\section{ Result and Discussion} 
\label{sec:r&d}
The H$-$H separation is one of the most important factors in identifying 
the potential candidate for hydrogen storage, because if the H-H separation
is small one can accommodate more H within a small region. From this point 
of view,Th$_2$AlH$_4$ may be considered as a potential candidate for 
storing H. To the best of our knowledge no theoretical or experimental
attempts have been made to study cohesive
properties like heat of formation (${\Delta H}$), cohesive energy (${E_{coh}}$), bulk
modulus ${(B_0)}$ and its pressure derivative ${(B_{0}^{\prime})}$ for this compound.
Hence, this is the first theoretical attempt to study the ground state 
properties and bonding in this compound.

\subsection{Structural optimization from total energy studies}

In order to analyze the effect of hydrogenation on the crystal structure of
Th$_2$Al and to verify the discrepancy between the experimentally observed H-H
separation, we have optimized the structural parameters for Th$_2$Al and
Th$_2$AlH$_4$. For this purpose, first we have relaxed the atomic positions
globally using the force-minimization technique, by keeping experimental $c/a$ and
cell volume ${(V_0)}$ fixed to experimental values. Then the theoretical equilibrium 
volume is determined by
fixing optimized atomic positions and experimental $c/a$, and varying the cell
volume by $\pm 10$ \% of ${V_0}$. Finally the optimized $c/a$ ratio is obtained by
a $\pm 2$ \% variation in $c/a$ ratio (in steps of 0.005), while keeping the
theoretical equilibrium volume fixed. It is important to note that
experimentally observed lattice parameters are almost same, while the atomic
position of H alone differs between the two experimental results
(according to Bergsma {\em et al.}\cite{bergsma61} H coordinates are 0.368, 0.868,
0.137 and S{\o}rby {\em et al.}\cite{sorby20} give 0.3707 , 0.8707, 0.1512). The total
energy ${vs}$. cell volume and $c/a$ ratio curves for Th$_2$Al and Th$_2$AlH$_4$
are shown in Figs.\,\ref{fig:vol1} and \,\ref{fig:vol2}, respectively. From
these illustrations it is clear that the equilibrium structural parameters obtained
from our theoretical calculations are in very good agreement with those
obtained recently by PND measurements.\cite{sorby20} 
\par
The optimized atomic positions along with the corresponding
experimental values are given in Table\,\ref{table:atompos}. 
Table\,\ref{table:interat} gives calculated lattice
parameters and interatomic distances, along with corresponding experimental values
for both Th$_2$Al and Th$_2$AlH$_4$. The theoretically estimated 
equilibrium volume is underestimated by
0.27\% for Th$_2$Al and 1.8 \% for Th$_2$AlH$_4$. The underestimation of bond
length in the present study is partly due to the limitation of local density
approximation used in the calculations and also neglect of the zero-point motion
and thermal expansions.
The difference between the experimental values may be due to the poor resolution
of the earlier(1961) PND data\cite{bergsma61}.

\subsection{Cohesive properties}

The method of calculation for cohesive properties for intermetallic
compounds are well described in 
Refs.\onlinecite{ravindran96,ravi99,vaj01} 
The cohesive energy is a measure of the force that binds atoms together in the solid
state. The cohesive energy of a system is defined as the sum of the total energy of 
constituent atoms at infinite separation minus the total energy of the particular
system. This is a fundamental property which has long been the subject of
theoretical approaches. The chemical bonding in intermetallic
compounds is a mixture of covalent, ionic and metallic bonding and therefore the
cohesive energy can not be determined reliably from simple models. Thus, first
principle calculations based on DFT have become a
useful tool to determine the cohesive energy of solids. For the study of
phase equilibrium the cohesive energy is more descriptive than the total energy,
since the latter includes a large contribution from electronic states that do
not play a role in bonding. From our cohesive energy calculations we get
${E_{coh}}$ = 0.15, 0.185 eV/atom for Th$_2$Al and Th$_2$AlH$_4$ respectively,
indicating that hydrogenation enhances the bond strength in Th$_2$Al.
\par
The formation energy ${(\Delta H)}$ is introduced in order to facilitate a
comparison of system stability. ${\Delta H}$ is defined as the total energy difference
between the compound and weighted sum of the corresponding total energy of the
constituents. For the ${\Delta H}$ calculations, we used the total energy value of 
\-2.320 Ry for the hydrogen
molecule which was calculated with the von Barth-Hedin exchange-correlation
potential.\cite{gunnarsson76} ${\Delta H}$ provides information about the stability
of Th$_2$Al towards hydrogenation.
The calculated ${\Delta H}$ values for La-Ni based hydrides,\cite{nakamura98} were almost
double the experimental\cite{murray81,bowerman79,hubbard83}
${\Delta H}$. As the LMTO-ASA method was used in this study, this discrepancy 
is expected because the internal relaxation of the atoms
was not taken into account and the interstitial potential is not well
represented in LMTO-ASA method. Therefore, ${\Delta H}$ calculated by using the
full-potential method should be more reliable. Our calculated 
values for ${E_{coh}}$ and ${\Delta H}$ are given in Table\,\ref{table:gro}. 
Since ${(\Delta H)}$ is more negative and ${E_{coh}}$ is higher for Th$_2$AlH$_4$ than
for Th$_2$Al, we can conclude that Th$_2$AlH$_4$ is more stable than Th$_2$Al.
However, no experimental ${\Delta H}$ values for
Th$_2$Al and Th$_2$AlH$_4$,  are available, but we
note that our calculated ${\Delta H}$
is close to the experimentally observed values of other Th-based hydrides,
like ThNi$_5$H$_4$, ThCo$_5$H$_4$ and ThFe$_5$H$_4$ all having the
${\Delta H}$ value of $-$36.63 kJ/(mol-H).\cite{vanmal73}
\par
The bulk modulus of Th$_2$Al and Th$_2$AlH$_4$ was obtained by 
self-consistent total energy calculations for 8 different volumes within the range of 
$V/V_0$ from 0.75 to 1.10 (see Fig.\,\ref{fig:vol1}). From the derivative of total
energy with respect to the volume, the calculated bulk modulus 
for Th$_2$Al is 93.42\,GPa and 
for Th$_2$AlH$_4$ is 111.36\,GPa. The corresponding pressure derivative
of the bulk modulus ${(B_{0}^{\prime})}$
are 3.43 and 3.51, respectively. The enhancement in ${B_0}$ value in the
hydrogenated phase
indicates that hydrogen plays an important role in bonding behavior of Th$_2$AlH$_4$.
In particular, the hydrogenation enhances the bond strength, and hence the change
in volume with hydrostatic pressure decreases with hydrogenation. This conclusion
is consistent with the observation made from our calculated heat of formation 
and cohesive energy for Th$_2$Al and Th$_2$AlH$_4$. 

\subsection{Anisotropic behavior}
For compounds which maintains the basic structural frame work,
the occupancy of hydrogen in interstitial sites is 
determined by its chemical environment
(different chemical affinity for the elements in the coordination sphere also
results in different occupancy). Although the H atom is small and becomes
even smaller by chemical bonding to the host, it may deform and stress the
host metal considerably depending upon the chemical environment. 
Lattice expansion usually of the order of 5 to 30\%,
often anisotropic, results from hydride formation. The maximum volume expansion observed for 
CeRu$_2$ to CeRu$_2$D$_5$ (37\%) is due to a hydrogen induced electron transition as shown by XPS 
measurements.\cite{osterwalder85}
A lattice contraction upon hydrogenation has so far only been observed for ThNi$_2$ to ThNi$_2D_2$ 
(\-2.2\%). For most hydrides formed from intermetallic compounds the crystal structure
usually changes with a loss of symmetry.\cite{schlapbach94} In general the symmetry  
decreases as a function of hydrogen content and increases as a function of temperature. 
However, on hydrogenation of Th$_2$Al the symmetry remains unchanged. 
\par
The volume expansion during hydrogenation of Th$_2$Al is 12.47\% ${(\Delta V}$/H
atom is 10.32{\AA}$^3$). This volume expansion is strongly anisotropic and proceeds
predominantly perpendicular to the basal plane of the tetragonal unit cell;
${{ \Delta }a/a}$ = 0.026\%, ${{ \Delta }c/c}$ = 12.41\%. This indicates a relatively
flexible atomic arrangement in the [001] direction. In spite of the isostructurality 
between Th$_2$Al, Zr$_2$Fe (hydrated: Zr$_2$FeH$_5$)\cite{yartys98}
and Zr$_2$Co (hydrated: Zr$_2$CoH$_{4.82}$)\cite{bonhomme93} the 
latter two exhibit a quite opposite
anisotropic behavior in that the the unit cell expands exclusively along the basal
plane.  
The $c/a$ ratio plays an important role for the structural properties of intermetallic compounds
including metal hydrides. For example, in the case of Zr$_2$Fe\cite{yartys98}, 
Zr$_2$Co\cite{bonhomme93},
Zr$_2$Ni\cite{chikdene89,elcombe99} and Th$_2$Al $c/a$  is 0.878, 0.867, 0.812 and 0.7695
respectively, and for the corresponding hydrides Zr$_2$FeH$_5$,\cite{chikdene89} 
Zr$_2$CoH$_{4.82}$,\cite{chikdene89}  Zr$_2$NiH$_{4.74}$\cite{chikdene89} and Th$_2$AlH$_4$
$c/a$ is 0.810, 0.815, 0.833 and 0.8543 respectively (see Fig.\,\ref{fig:cba}). 
The increase in $c/a$ for Zr$_2$CoH$_{4.82}$ and Zr$_2$FeH$_5$ compared
with the corresponding unhydrated parents is smaller than that for other compounds. 
On hydrogenation, the increase in $c/a$ ratio for Th$_2$Al is
considerably larger than for Zr$_2$Ni, which may be the reason
why the former retains the symmetry on hydrogenation. Our calculations describe well the
anisotropic changes in the crystal structure on hydrogenation of Th$_2$Al (see Table\,\ref{table:interat}). 
The $c/a$ ratio increases almost linearly (Fig.\,\ref{fig:cba}) on going from Zr$_2$Fe
to Th$_2$Al whereas the corresponding hydrides show the
opposite behavior. Hence, it appears that the systematic variation in $c/a$
plays a major role in deciding the crystal structure for the
CuAl$_2$-type hydrides. When  $c/a$ $<$ 0.825 the symmetry is changed
from $I4/mcm$ to $P4/ncc$ on hydrogenation, whereas when  $c/a$ $>$
0.825 the crystal symmetry is apparently not affected. 

\subsection{Electronic structure}
In order to understand the changes in the electronic bands on hydrogenation of
Th$_2$Al we  show the energy band structure for Th$_2$Al and 
Th$_2$AlH$_4$ in Fig.\,\ref{fig:bnd}a and $b$ respectively. The illustration clearly indicates
that inclusion of H in the Th$_2$Al matrix has a noticeable impact on the 
band structure, mainly in the valence band. The two lowest lying broad 
bands in Fig.\,\ref{fig:bnd}a originate from Al 3$s$ electrons. As the
unit cell contains two formula units, eight electrons are additionally introduced 
when Th$_2$AlH$_4$ is formed from Th$_2$Al.
These electrons form four additional bands (Fig.\,\ref{fig:bnd}b), 
a large deformation of the band structure is introduced by the of hydrogen
in the Th$_2$Al lattice. These bands become localized, the lowest lying 
energy band is moved from $-$7.34 to$-$9.20\,eV, and, the character of the 
latter band is changed from Al 3$s$ to
H 1$s$ character. The Al $3s$ bands are located in a wide energy range from
$-$2.8 to $-$7.34\,eV in Th$_2$Al and are in a narrow energy range from $-$2.5 
to $-$4.2\,eV in the hydride. The drastic change in the Al bands on hydrogenation
of Th$_2$Al is due to the electron transfer from H to Al and this is discussed
further in Sec.\,\ref{sec:bond}. The H $s$ bands are found in the energy
range from $-$2\,eV to the bottom of the valence band. Their contribution at $E_{F}$ is
negligibly small indicating the formation of localized bands. The bands
at  E$_F$ is dominated by the Al $3p$ and Th $6d$ electrons in both Th$_2$Al
and Th$_2$AlH$_4$. Owing to the creation of the pseudogap feature 
near ${E_F}$, the contribution of the  Al $3p$ electrons
to the bands at ${E_F}$ level are significantly reduced by
the hydrogenation of Th$_2$Al.
    
\subsection{Nature of Chemical Bonding}
\label{sec:bond}
\subsubsection{ Density of state}

In order to obtain a deeper insight into the changes in chemical bonding behavior
on hydrogenation of Th$_2$Al we  give the angular-momentum and 
site-decomposed DOS for Th$_2$Al and Th$_2$AlH$_4$ in Fig.\,\ref{fig:dos}. 
DOS features for Th$_2$Al and Th$_2$AlH$_4$ show close similarity. 
Both exhibit metallic character since there is finite DOS at $E_F$.
From the DOS histogram we see that $E_F$ is systematically shifted 
towards higher energy in Th$_2$AlH$_4$. This is due
to the increase in the number of valence electrons when Th$_2$Al is hydrogenated.
DOS for both Th$_2$Al and Th$_2$AlH$_4$ lie mainly in four energy regions (a) the
lowest region around $-$20\,eV stems mainly from localized or tightly bound
Th $6p$ states, (b) the region from $-$9.25 to $-$2.5\,eV originates from bonding 
of H $1s$, Al $3p$ and Th $6d$  (Th-$6d$ and Al-$3p$ states in Th$_2$Al), 
(c) the region from $-$2.5 to 0\,eV comes from bonding states of Al $3p$
and Th $6d$ and (d) the energy region just above $E_F$ (0 to 3.5\,eV)
are dominated by unoccupied Th $4f$ states. 
\par
The semi-core Th $6p$ states are well localized and naturally their effect on bonding is very
small. On comparing the Th $6p$ DOS of Th$_2$Al and Th$_2$AlH$_4$, it is
seen that the width is significantly reduced in Th$_2$AlH$_4$ owing to the lattice 
expansion and the inclusion of additional energy levels below $E_F$. 
In the valence band region, the band width and DOS are larger for Th$_2$AlH$_4$
than for Th$_2$Al. Hydrogenation enhances interaction between
neighboring atoms and thereby increases
the overlap of orbitals and in turn results in the enlarged  valence band 
width in Th$_2$AlH$_4$. In particular, the strong hybridization between
Th $6d$ and H $1s$ states increases the valence band width from 7.1\,eV
in Th$_2$Al to 8.4\,eV in Th$_2$AlH$_4$. Th $6d$, Al $3p$ and H $1s$ states are
energetically degenerate in the valence band region indicating a possibility 
of covalent Th-H, Th-Al and Al-H bonds. However, the spatial separation
Th-Al (3.22\,{\AA}) and Al-H (3.02\,{\AA}) is larger than Th-H (2.26\,{\AA}). 
Therefore, covalent bonds between the former part is small whereas there is a significant
covalent contribution between Th and H. In conformity with this the 
COHP and charge density analyses show directional bonding between Th 
and H (see Sec.~\ref{sec:charge} and ~\ref{cohp}). 
The accommodation of H in the interstitial position between
Th and Al creates new bonding states between Th and H. This
also enhances Th-Al distance around 2.2\% compared with that in Th$_2$Al.
The consequence of this enhancement is that the Al DOS at the valence band region becomes
narrow and the splitting between the Al 3$s$ and 3$p$ states is 
almost doubled (see  Fig.\ref{fig:dos}). The finite DOS at $E_F$ which 
gives the metallic character of Th$_2$Al and Th$_2$AlH$_4$
comes from Th$d$ states in addition to some states of Al $p$ character.
\par
Another interesting feature of the total DOS of Th$_2$AlH$_4$ is presence of a deep
valley around $E_F$ which is termed as a pseudogap. Pseudogap features are
formed not only in crystalline solids\cite{xu90} but occur 
also in amorphous phases\cite{pasturel85} 
and quasicrystals.\cite{phillips92} Two mechanisms have been proposed for occurrence of
pseudo gap in binary alloys, one attributed to ionic features and the other to the effect of 
hybridization. Although the electronegativity differences between Th, Al and H
are noticeable, they are not large enough to explain the pseudogap in Th$_2$AlH$_4$. 
Hence hybridization must be the cause for creation of the pseudogap in Th$_2$AlH$_4$. 
There is a correlation  
between the occurrence of pseudogaps and structural stability\cite{ravi97}, that materials
which possess pseudogaps in the vicinity of $E_F$ usually have higher stability.
This may be the reason for the higher value of $\Delta H$ in Th$_2$AlH$_4$ than
in Th$_2$Al(Table\,\ref{table:gro}). 

\subsubsection{Charge density}
\label{sec:charge}
The analysis of the bonding between the constituents will give better understanding
about the anisotropic changes in the structural parameters on hydrogenation of Th$_2$Al. 
Fig.\,\ref{fig:ch1}) shows the calculated valence charge density 
(obtained directly from the self-consistent calculation) within  
$ab$ and $ac$ planes for Th$_2$AlH$_4$. 
Th, Al and H atoms are confined to layers along $c$, Th and Al
being situated in alternating metal layers with hydrogen in between, hence
establishing a sequence of Th-H-Al-H-Th-H-Al-H-Th layers (see Fig.\,\ref{fig:cry}). 
The H atoms are arranged in a chain like manner within the $ab$ plane as also evident from
Fig.\,\ref{fig:ch1}b. It is interesting to note that the nature of H-H bonding is quite 
different along $a$ and $c$. Although the H-H 
distance is 2.34\,\AA within the basal plane and 1.95\,\AA perpendicular to the
basal plane, the bonding  between the H atoms is not totally dominated by 
the latter interactions. In fact the 
COHP analysis (sec.~\ref{cohp}) shows that the covalent H-H interaction within the
$ab$ plane is larger than within $ac$ plane. The electron
distribution between Al and H suggests ionic bonding between them, in line with their
electronegativity difference of 0.7. In conformity with this the integrated
charge inside the Al sphere is around 0.59 (0.8 according to the TBLMTO method) 
electrons larger in Th$_2$AlH$_4$ than in Th$_2$Al.  
\par
The bonding between Th and H is predominantly covalent as evidenced by the 
finite charge between these atoms. The H-$s$ electrons
are tightly bound to the Th-$d$ states, and Th-H arrangement forms a  
H-Th-H dumb-bell pattern. Now we will try to obtain a possible explanation 
for the short H-H distance within $ac$ plane of Th$_2$AlH$_4$ from the 
charge density analysis. The strong covalent interaction between Th and 
H in the $ac$ plane(see Fig.\,\ref{fig:ch1}b) and the dumb-belled pattern tend to draw    
the electrons of H towards Th leaving only a small amount
of electrons between the H along $c$ to repul each other. The main reason
for this short H-H distance is then reduced repulsion rather than 
bonding interaction between them.

\subsubsection{COHP}
\label{cohp}
COHP is an extremely useful tool to analyze covalent bonding interaction 
between atoms, the simplest approach being to investigate 
complete COHP between the atoms concerned, taking all valence orbitals into account. 
COHP between Th-H, Th-Al, Al-H and H-H in Th$_2$AlH$_4$ are given in Fig.\,\ref{fig:cohp}.
 
\par
Owing to the very different interatomic distances between the H atoms 
in the $ab$ and $ac$ planes, special attention is paid to COHP in these planes. 
Both bonding and antibonding states are present almost equally in
the VB region indicating that covalent interaction between 
the H atoms is not participating significantly to the stability of  
Th$_2$AlH$_4$. On the other hand, the bonding states are present in the whole
VB region in COHP of Th-Al and Th-H indicating that covalent 
interaction between these pairs is contributing to structural
stability. The presence of the large bonding states in the VB region of COHP
for Th-H is the main reason for large value of  heat of formation for
Th$_2$AlH$_4$  compared with Th$_2$Al. In order to quantify the covalent
interaction between constituents of Th$_2$AlH$_4$ we have integrated the COHP curves  
up to $E_F$ for Th-Al, Th-H and Al-H giving $-$0.778, $-$1.244 and $-$0.072, 
respectively. Owing to the presence of both bonding
and antibonding states below $E_F$ in COHP the integrated
value for H-H becomes negligibly small ($-$0.086 and $-$0.011 within the $ac$ and 
$ab$ plane, respectively, but as the integrated value of bonding states alone is
$-$0.571 and $-$0.136 respectively, the bonding H-H interaction is quite
different in the two planes). Hence, one
can conclude that the bond strength between the constituents of Th$_2$AlH$_4$
decrease in the order Th-H $>$  Th-Al $>$ Al-H $>$ H-H.

\par
The experimental\cite{bergsma61,sorby20} and theoretical studies show highly
anisotropic changes in the lattice expansion on hydrogenation of Th$_2$Al.
According to the crystal structure of Th$_2$Al the interatomic distance between
the interstitial regions where one can accommodate H in the $ab$ plane is 2.4\,{\AA}.
Hence, there is a large flexible space for accommodation of the H atoms in this plane 
without the need to expand the lattice. In contrast, the interatomic distance
between the interstitial regions in the $ac$ plane is only 1.65\,{\AA}. 
So, large expansion of the lattice along $c$ is necessary
to accommodate H within the $ac$ plane. As a result, even with a 
short H-H separation of 1.95\,\AA, a volume expansion of 12.41\% 
is needed when Th$_2$AlH$_4$is formed from Th$_2$Al. The experimental observation
of 0.105\% lattice expansion along $a$ and 12.15\% along $c$ is found to be in excellent
agreement with theoretically obtained value of 0.03  and 12.41\%, respectively.  
 
\section{Conclusion}
\label{sec:s&c}

This study reports a detailed investigation on the electronic structure, 
bonding nature and ground state properties of Th$_2$Al and Th$_2$AlH$_4$ 
using first-principle method. The following important conclusions are obtained. \\

\noindent 
1) Th$_2$Al and Th$_2$AlH$_4$ are formed in
 the CuAl$_2$-type
crystal structure, the optimized atomic positions and lattice parameters  
 are in very good agreement with recent experimental results.\\
\noindent 
2) Structural optimization gave a shortest 
 H-H separation of 1.95\,\AA, which is  
 close to the recent experimental value of 1.97\,\AA. \\
\noindent
3) We observed a highly anisotropic volume expansion of 12.47\% of the
 Th$_2$Al matrix on hydrogenation to Th$_2$AlH$_4$, of which  
  99.76\% volume expansion occurs perpendicular to the basal plane and neglible change
  along the basal plane.\\
\noindent
4) The large difference in interatomic distance between the interstitial regions
within the $ab$ and $ac$ planes and 
the strong covalent interaction between Th and H along $c$ keeps the
H atoms close together in the $c$ direction. This is
the main reasons for highly anisotropic volume
expansion on hydrogenation in Th$_2$Al. \\
\noindent
5) Charge density and COHP analysis revealed that the Th-H bonds are stronger
than the H-H bonds and other localized bonds in this structure. The formation
of strongly bonded ThH$_2$ subunits in Th$_2$AlH$_4$ makes repulsive 
interaction between the H atoms smaller along $c$ and this is the precise reason 
for the violation of 2\,\AA rule.\\
\noindent
6) There is a correlation between $c/a$ and the structural 
stability of hydrated CuAl$_2$-type phases. For phases with
$c/a < $ 0.825 the symmetry changes from
$I_4/mcm$ to $P_4/ncc$ on hydrogenation, whereas for $c/a > $ 0.825 the crystal
symmetry is not affected on hydrogenation.\\
\noindent 
7) Density of states and bandstructure studies show that Th$_2$Al and Th$_2$AlH$_4$ are 
having non vanishing N(E$_F$), resulting in metallic character.   
The cohesive energy analysis show that, Th$_2$AlH$_4$ is more stable than
Th$_2$Al.   

\acknowledgements 
P.V. gratefully acknowledges Prof. Karlheinz Schwarz, Prof. Peter Blaha, Prof. O.K. Andersen and 
Prof. O. Jepsen for supplying computer codes used in this study. The authors also acknowledge
Dr. Florent Boucher for useful discussion on COHP. This work has received support from 
The Research Council of Norway (Programme for Supercomputing) through a grant of 
computing time.
 
%%%%%%%%%%%%%%%%%%%%%%%%% Table 1 %%%%%%%%%%%%%%%%%%%%%%%%%%%%%%%%%%
\begin{table}
\caption{ Atomic position of Th$_2$Al and Th$_2$AlH$_4$}  
\begin{tabular}{|l|c c c |c c c|}   
 & \multicolumn{3}{c|} {Th$_2$ Al~~} & \multicolumn{3}{c|}{Th$_2$AlH$_4$} \\ \cline{2-4}\cline{4-7}  
            & $x$      & $y$      & $z$      &$x$       & $y$      & $z$   \\ \hline
Th~~Theory                  & 0.1583 & 0.6583 & 0.0000 & 0.1632 & 0.6632 & 0.0000 \\
~~~~~Exp.\cite{pearson86}    & 0.1588 & 0.6588 & 0.0000 &  --    & --     & --      \\
~~~~~Exp.\cite{sorby20}    &   --   & --     & --      & 0.1656 & 0.6656 & 0.0000  \\
~~~~~Exp.\cite{bergsma61}  &        &        &        & 0.162  & 0.662  & 0.0000 \\ \hline
Al~~Theory                & 0.0    & 0.0    & 0.25   & 0.0    & 0.0    & 0.25   \\
~~~~~Exp.\cite{pearson86}  & 0.0    & 0.0   & 0.25    & --     & --     & --    \\
~~~~ Exp.\cite{sorby20}    &  --    &  --    &  --    & 0.0    & 0.0    & 0.25   \\
~~~~~Exp.\cite{bergsma61}  &   --   &  --    &  --    & 0.0    & 0.0    & 0.25   \\  \hline
H~~~Theory &  --    & --     & --     & 0.3705 & 0.8705 & 0.1512 \\
~~~~ Exp.\cite{sorby20}    &  --    & --     & --     & 0.3707 & 0.8707 & 0.1512  \\
~~~~~Exp.\cite{bergsma61}  &  --    & --     & --     & 0.368  & 0.868  & 0.137   \\  
\end{tabular}
\label{table:atompos}
\end{table} 
%%%%%%%%%%%%%%%%%%%%%%%%% Table 1 end %%%%%%%%%%%%%%%%%%%%%%%%%%%%%%%%%%
%%%%%%%%%%%%%%%%%%%%%%%%% Table 2 %%%%%%%%%%%%%%%%%%%%%%%%%%%%%%%%%%
\begin{table}
\caption{ Lattice parameters and inter atomic distances  of Th$_2$Al
          and Th$_2$AlH$_4$(all values are in \,\AA).}
\begin{tabular}{|l|r|r|r|r|r|}
& \multicolumn{2}{c|} {Th$_2$ Al~~} & \multicolumn{3}{c|}{Th$_2$AlH$_4$} \\ \cline{2-3}\cline{3-6}
&Theory  &  Exp.\cite{pearson86}  & Theory  & Exp.\cite{sorby20}  & Exp.\cite{bergsma61}  \\  \hline
$a$      &  7.602 & 7.618 & 7.604 & 7.626 & 7.629 \\
$c$      & c = 5.723 & 5.862 & 6.433 & 6.515 & 6.517 \\
$\it{c/a}$    & 0.753 & 0.769 &  0.846 & 0.854 & 0.854 \\
Th-H & -- & -- & 2.273 & 2.305 & 2.387 \\
Th-Al & 3.199 & 3.219 & 3.269 & 3.278 & 3.291 \\
Th-Th & 3.403 & 3.421 & 3.509 & 3.571 & 3.495 \\
Al-H  & -- & -- & 3.051 & 3.061 & 3.072  \\
Al-Al & 2.861 & 2.931 & 3.216 & 3.257 & 3.258  \\
H-H ($ac$-plane) & -- & -- & 1.945 & 1.971 & 1.790 \\
H-H ($ab$-plane) & -- & -- & 2.344 & 2.305 & 2.495 \\
\end{tabular}
\label{table:interat}
\end{table}
%%%%%%%%%%%%%%%%%%%%%%%%% Table 2 end %%%%%%%%%%%%%%%%%%%%%%%%%%%%%%%%%%

%%%%%%%%%%%%%%%%%%%%%%%%% Table 3 %%%%%%%%%%%%%%%%%%%%%%%%%%%%%%%%%%
\begin{table}
\caption{ Ground state properties of Th$_2$Al and Th$_2$AlH$_4$}
\begin{tabular}{|l|c|c|}  
Compound & Th$_2$Al & Th$_2$AlH$_4$ \\ \hline 
$-$ $\Delta H$ ( in kJ mol$^{-1}$)  &  18.35   & 29.50    \\
 $E_{coh}$(eV/atom)	& 0.15  & 0.185   \\
 $N(E_F)$(states/Ry-cell)	& 58.42   & 41.13   \\
$B_0$(GPa)	&   93.42 & 111.36   \\
$B_{0}^{\prime}$	&   3.41& 3.48   
\end{tabular}
\label{table:gro}
\end{table}
%%%%%%%%%%%%%%%%%%%%%%%%% Table 3 end %%%%%%%%%%%%%%%%%%%%%%%%%%%%%%%%%%

%%%%%%%%%%%%%%%%%%%%%%%%% FIG.1 %%%%%%%%%%%%%%%%%%%%%%%%%%%%%%%%%%
\begin{figure}
\caption{The crystal structure of Th$_2$AlH$_4$. Five Th in face sharing
tetrahedral configuration surrounding two hydrogen.  
Legends to the different kinds of atoms are given on the illustration.}
\label{fig:cry}
\end{figure}
%%%%%%%%%%%%%%%%%%%%%%%%% FIG.2 %%%%%%%%%%%%%%%%%%%%%%%%%%%%%%%%%%
\begin{figure}
\caption{ Total energy (a) ${vs.}$ $c/a$  and (b)   
${vs.}$  unit cell volume for Th$_2$Al where $\Delta E$=$E$+106632. } 
\label{fig:vol1}
\end{figure}
%%%%%%%%%%%%%%%%%%%%%%%%% FIG.3 %%%%%%%%%%%%%%%%%%%%%%%%%%%%%%%%%%
\begin{figure}
\caption{ The total energy (a) ${vs.}$ $c/a$ and  (b) 
 ${vs.}$ unit cell volume  for Th$_2$AlH$_4$ where $\Delta E$=$E$+106636.}
\label{fig:vol2}
\end{figure}
%%%%%%%%%%%%%%%%%%%%%%%%% FIG.4 %%%%%%%%%%%%%%%%%%%%%%%%%%%%%%%%%%
\begin{figure}
\caption{ $c/a$ for CuAl$_2$-type phases and their corresponding hydrides. }
\label{fig:cba}
\end{figure}
%%%%%%%%%%%%%%%%%%%%%%%%% FIG.5 %%%%%%%%%%%%%%%%%%%%%%%%%%%%%%%%%%
\begin{figure}
\caption{Electronic band structure of (a) Th$_2$Al and (b) Th$_2$AlH$_4$. The
Fermi level is set to zero.}
\label{fig:bnd}
\end{figure}
%%%%%%%%%%%%%%%%%%%%%%%%% FIG.6 %%%%%%%%%%%%%%%%%%%%%%%%%%%%%%%%%%
\begin{figure}
\caption{ Total, site and orbital projected density of states for (a) Th$_2$Al and 
        (b) Th$_2$AlH$_4$. }
\label{fig:dos}
\end{figure}
%%%%%%%%%%%%%%%%%%%%%%%%% FIG.7 %%%%%%%%%%%%%%%%%%%%%%%%%%%%%%%%%%
\begin{figure}
\caption{ Valence electron charge density plot for Th$_2$AlH$_4$ 
in the $ab$ plane 
with 40 contours drawn between 0 and 0.25 electrons/a.u.$^3$ }
\label{fig:ch1}
\end{figure}
%%%%%%%%%%%%%%%%%%%%%%%%% FIG.8 %%%%%%%%%%%%%%%%%%%%%%%%%%%%%%%%%%
\begin{figure}
\caption{Valence electron charge density plot between the Th and H atoms for Th$_2$AH$_4$  
in the $ac$ 
plane with 40 contours drawn between 0 and 0.25 electrons/a.u$^3.$ }
\label{fig:ch2}
\end{figure}
%%%%%%%%%%%%%%%%%%%%%%%%% fig.9 %%%%%%%%%%%%%%%%%%%%%%%%%%%%%%%%%%
\begin{figure}
\caption{COHP of Th$_2$AlH$_4$, depicting the contributions from
Th-Al, Th-H, Al-H and H-H interactions. The COHP for H atoms in the $ab$ plane
and $ac$ plane are given in solid line and dotted lines respectively.}
\label{fig:cohp}
\end{figure}
%%%%%%%%%%%%%%%%%%%%%%%%% end FIG. %%%%%%%%%%%%%%%%%%%%%%%%%%%%%%%%%% 

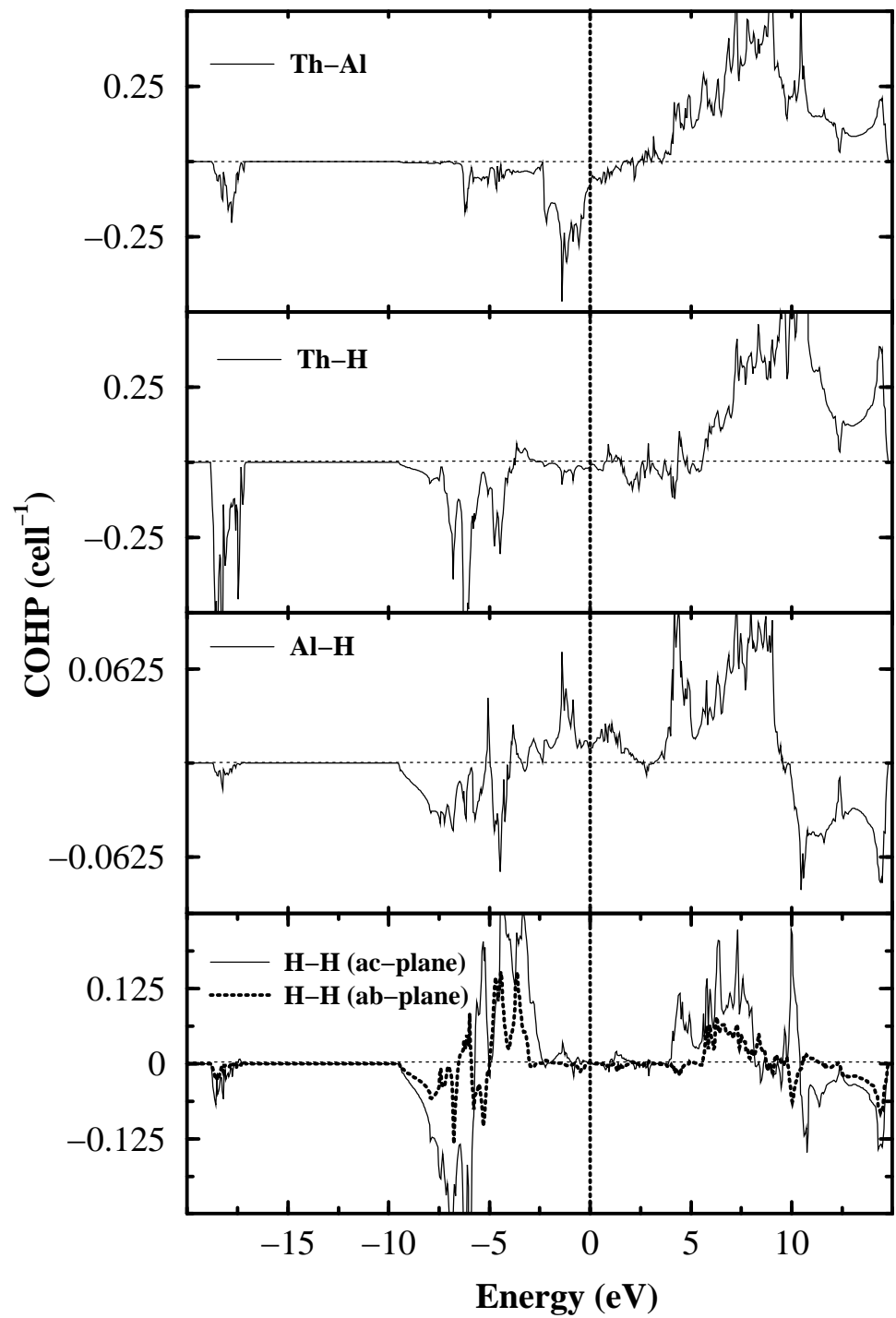
\end{document}